\documentstyle[12pt]{article}
 \begin{document}
 \baselineskip=12pt
\newcommand{\ekq}{e^{2\pi i\frac{k}{q}}}
 \newcommand{\rf}{Ramanujan - Fourier~}
\newcommand{\ekqn}{e^{2\pi i\frac{k}{q}n}}
 \newcommand{\ekqm}{e^{-2\pi i\frac{k}{q}}}
 \newcommand{\ekqmn}{e^{-2\pi i\frac{k}{q}n}}
 \newcommand{\wk}{Wiener - Khintchine formula~}
 \newcommand{\ekqp}{e^{2\pi i\frac{k'}{q'}}}
 \newcommand{\di}{\displaystyle}

 \begin{center}
 {\Large \bf Rota meets Ramanujan: Probabilistic interpretation of Ramanujan - Fourier series}\\
 \vspace{1cm}
 {H. Gopalkrishna Gadiyar and R.Padma\\
AU-KBC Research Centre, M.I.T. Campus, Anna University\\
Chromepet, Chennai 600 044 India\\
E-mail: gadiyar@au-kbc.org, padma@au-kbc.org}
 \end{center}
 \vspace{2cm}
 \begin{center}
 {\bf Abstract}
 \end{center}
 In this paper the ideas of Rota and Ramanujan are shown to be central to understanding problems in additive number theory. The circle and sieve methods are two different facets of the same theme of interplay between probability and Fourier series used to great advantage by Wiener in engineering.
 \vspace{1cm}

Norbert Wiener forged a powerful tool for electrical engineers by combining two distinct branches of probability and Fourier series. Recently the authors in [1] have shown that the twin prime problem is related to the Wiener-Khintchine formula for Ramanujan - Fourier expansion for a relative of the von Mangoldt function. Planat[2] has extensively developed applications of Ramanujan - Fourier series in practical settings. The next natural question is: Is there a probabilistic interpretation of \rf series? To our surprise we found that there are two distinct streams of thought one due to Rota which is combinatorial and cast in the modern, abstract language of characters and the other is the historically older argument of Ramanujan in the  classical, concrete language of Fourier series.

We would like to give the punch line right away and then give a brief summary of the view points of Rota and Ramanujan. Rota considers the group $C_{\infty}$ of rational numbers modulo 1 and crucially bases his arguments summarized in the next section on $C_{\infty}^*$ the group of characters of $C_{\infty}$. Ramanujan Fourier series are
$$
 a(n) ~=~ \sum_{q=1}^\infty a_q c_q(n) \, ,
$$
where ${\displaystyle {c_q(n) ~=~ \sum_{\stackrel{k=1}{(k,q)=1}}^q e^{2\pi i
 \frac{k}{q}n }}}$. {\it Both are using the same tool as $e^{2\pi i
 \frac{k}{q}n}$ of Ramanujan is a concrete realization of the characters of the profinite group of Rota}. It seems that Rota did not notice the Ramanujan connection. As Sieve methods [3] are based on probabilistic considerations, the circle and the sieve methods are linked through the ideas of Rota and Ramanujan.

\noindent Rota: Recently Rota gave a stochastic interpretation of the Riemann - zeta function [4],[5]. Consider $A$ a subset of positive integers $\cal N $, then the arithmetic density is defined as 
$$
dens(A) = \lim_{n \rightarrow \infty} \frac{1}{n} || A \cap \{ 1,2,...,n\} ||
$$ 
whenever the limit exists. It is immediately obvious that the dens($\cal N $)=1 and $dens(A_p)=\frac{1}{p}$ where $A_p$ is the set of multiples of $p$. For technical reasons (lack of countable additivity) the right way to go about it is to choose a number $s > 1$ define the measure of a positive integer $n$ to be $\frac{1}{n^s}$. Then it turns out that the measure of $\cal N $ is equal to $\di{\zeta (s) = \sum_{n=1}^\infty \frac{1}{n^s}}$
Hence a countably additive measure $P_s$ on the set $\cal N$ defined as
$$
P_s(A) = \frac{1}{\zeta (s)} \sum_{n \in A} \frac{1}{n^s}
$$
Further the fundamental property 
$$
P_s(A_p \cap A_q) = P_s(A_p) P_s(A_q) = \frac{1}{p^sq^s}
$$
can be checked and ${\di \lim_{s\rightarrow 1} P_s(A) = dens(A)}.$
That is, the arithmetic density though not a probability is the limit of probabilities. 

Rota then gives a combinatorial twist to the problem. He considers a cyclic group $C_r$ of order $r$. Every character $\chi$ of the group $C_r$ has a kernel which is a subgroup of $C_r$. Further every sequence $\chi_1 ,..., \chi_s$ of characters of $C_r$ has a joint kernel which is a subgroup of $C_r$. By a joint kernel of a sequence of characters we mean the intersection of their kernels. Suppose one were to choose a sequence of $s$ characters independently and randomly and ask the question: what is the probability that the joint kernel is a certain subgroup $C_n$ of $C_r$? It can be seen that with probability $\frac{1}{n}$ the kernel of a randomly chosen character will contain the subgroup $C_n$ as there are $r$ characters of the group $C_r$ and $\frac{r}{n}$ such characters will vanish on $C_n$. So the probability of the joint kernel will contain $C_n$ is $\frac{1}{n^s}$. Let $P_{C_n}$ denote the probability that the joint kernel of characters shall be $C_n$. It follows that ${\di
\frac{1}{n^s} = \sum_{n |d|r} P_{C_d}}$
This is based on the fact that the partially ordered set of subgroups of a cyclic group $C_r$ and the partially ordered set of the divisors of r are isomorphic. Next using Moebius inversion and change of variable  $d= nj$ it follows that 
$$
P_{C_n} = \sum_{n|d|r} \mu(\frac{d}{n}) \frac{1}{d^s}
$$
$$
P_{C_n} = \frac{1}{n^s} \sum_j \mu(j) \frac{1}{j^s}
$$
The variable $j$ ranges over a subset of divisors of $r$. If the sum ranged over all positive integers $j$ we would get the probabilistic interpretation. 
$$
\frac{1}{n^s} \frac{1}{\zeta(s)}
$$
This is done by replacing the finite cyclic group $C_n$ by a profinite cyclic group. Take the group $C_{\infty}$ of rational numbers modulo 1. For every positive integer $n$ the group $C_{\infty}$ has unique finite subgroup $C_n$ of order $n$. The character group $C_{\infty}^*$ of $C_{\infty}$ is a compact group, and the Haar measure is a probability measure. The group $C_{\infty}^*$ is the desired profinite group. Mimicking the argument made earlier it can be seen that
$$
 \frac{1}{n^s}=\sum_{n |d} P_{C_d}
$$
and by Moebius inversion
$$
P_{C_n} = \sum_{n|d} \mu(\frac{d}{n}) \frac{1}{d^s} = \frac{1}{n^s} \frac{1}{\zeta(s)}
$$ 

\noindent Ramanujan, Hardy and Littlewood:  The first problem to be attacked by the circle method
 was the partition problem. As is well known, this reduces to understanding
 the generating function
 $$
 \sum_{n=0}^{\infty} p(n)\,x^n ~=~ \frac{1}{(1-x)(1-x^2)(1-x^3)......} \, .
 $$
 The key observation was
 that $p(n)$ could be written as
 $$
 \frac{1}{2\pi i} \int_C{\frac{1}{(1-z)(1-z^2)(1-z^3).....} z^{-n-1}dz} \, ,
 $$
 where $C$ is the unit circle $|z| ~=~1$. However, they observed that the
 unit circle is covered by infinity of singularities corresponding to the
 poles of the generating function at all the rational points on the unit circle.
 See [6]. Later Ramanujan showed by simple, yet ingenious methods that a wide
 range of arithmetical functions have
 Ramanujan - Fourier expansions, that is, an expansion of the form
 $$
 a(n) ~=~ \sum_{q=1}^\infty a_q c_q(n) \, ,
 $$
 where $ {\displaystyle {c_q(n) ~=~ \sum_{\stackrel{k=1}{(k,q)=1}}^q e^{2\pi i
 \frac{k}{q}n }}}$.
 R.D. Carmichael a little later gave the formula for the Fourier coefficients. He showed that
 $$
 a_q ~=~ \frac{1}{\phi (q)}  \lim_{X\rightarrow \infty} \frac{\sum_{n \le X} a(n)
 c_q(n)}{X} \, .
 $$
 Hardy stressed the multiplicative nature
 of $c_q(n)$ which he used to derive many formulae.
   
 \noindent {\bf References}
 \begin{description}
\item[1] H. Gopalkrishna Gadiyar, R. Padma {\it Ramanjuan-Fourier series, the Wiener-Khintchine formula and the distribution of prime pairs}, Physica A 269(1999) 503-510
\item[2] Planat, M, Rosu, H, Perrine, {\it Ramanujan sum for signal processing of low frequency noise} Submitted to Physical Review E
\item[3] Greaves, G. {\it Sieves in Number theory}, Springer - Verlag, 2001
\item[4] Alexander, K.S., Baclawski, K, Rota, G.C. {\it A stochastic interpretation of the Riemann seta function}, Proc. Natl. Acad. Sci. USA, Vol.90, 697-99, January 1993
\item[5] Rota, G.C. {\it Combinatorial snapshots}, The Mathematical Intelligencer, 21 (1999) No.2, 8-14
\item[6] Vaughan, R. C. {\it The Hardy - Littlewood method}, Cambridge Tracts in Mathematics 125, Cambridge University Press, 1981
 \end{description}
 \end{document}